\documentclass[
    aps,
    prd,
    reprint,
    nofootinbib,
    amsmath,
    amssymb,
    floatfix,
    ]{revtex4-2}

\usepackage{graphicx}

\usepackage{xcolor}
\definecolor{FJHGreen}{RGB}{28, 94, 46}
\definecolor{BurntOrange}{RGB}{202, 78, 16}
\definecolor{Gold}{RGB}{238, 155, 17}
\definecolor{GreyBlue}{RGB}{82, 117, 123}
\definecolor{WDMgrey}{RGB}{74, 74, 72}
\definecolor{DDgrey}{RGB}{83, 83, 83}

\usepackage[pdfpagemode={UseOutlines},bookmarks=true,bookmarksopen=true,
   bookmarksopenlevel=0,bookmarksnumbered=true,hypertexnames=false,
   colorlinks, citecolor={FJHGreen}, urlcolor={blue}, linkcolor={FJHGreen}, pdfstartview={FitV},unicode,breaklinks=true]{hyperref}

\usepackage[capitalise]{cleveref}

\begin{document}

\title{Photonic Freeze-In}

\author{Peter Cox}
\email{peter.cox@unimelb.edu.au}
\author{Matthew J. Dolan}
\email{matthew.dolan@unimelb.edu.au}
\author{Frederick J. Hiskens}
\email{frederick.hiskens@unimelb.edu.au}
\affiliation{ARC Centre of Excellence for Dark Matter Particle Physics, School of Physics, The University of Melbourne, Victoria 3010, Australia}

\begin{abstract}
We present a new scenario of freeze-in, where the dark matter is produced exclusively via the annihilation of photons. We study fermionic and scalar dark matter models with a focus on cosmological histories with low reheat temperatures. Photonic freeze-in can be probed via direct detection, dark matter production in SN1987A, and via the production of new electromagnetically charged states at the Large Hadron Collider. We briefly discuss UV-complete models that can realise this scenario.
\end{abstract}

\maketitle

\section{Introduction}

The freeze-in mechanism~\cite{Hall:2009bx} is a compelling solution to the origin of the cosmological dark matter relic density. In this paradigm, dark sector particles were not in equilibrium with the Standard Model (SM) in the early universe, usually due to an assumption of very weak interactions between the dark sector and the SM. Despite this, the correct relic density can be achieved through a gradual build-up of dark matter via the annihilations of SM particles. 

There are a multiplicity of simple models that realise the freeze-in mechanism, many of which utilise a tree-level portal interaction with a light vector or scalar mediator. In this work we present a minimal scenario with a loop-level origin, in which the dark matter couples via an effective interaction to the photon field-strength squared\footnote{This effective interaction can also produce dark matter via thermal freeze-out, a scenario known as Rayleigh dark matter~\cite{Weiner:2012cb,Frandsen:2012db,Kavanagh:2018xeh}.}, $F^{\mu\nu}F_{\mu\nu}$. This operator can be generated by integrating out heavy charged particles in a UV-complete model. This \textit{photonic freeze-in} scenario is a viable phenomenological possibility which has been underexplored. 

Photonic freeze-in can quite naturally be realised for any post-inflationary reheat temperature where the new heavy, electromagnetically charged states required for the UV completion are kinematically inaccessible.  Since the relevant effective operator is non-renormalisable, our scenario is an example of ultraviolet freeze-in~\cite{Elahi:2014fsa}. The correct dark matter abundance can be attained for dark matter masses as low as tens of keV, with lighter masses excluded by warm dark matter constraints.

We find that low reheat temperatures are particularly interesting phenomenologically. Models of freeze-in with a low reheat temperature have been constructed in which electrons~\cite{Bernal:2017kxu}, neutrinos~\cite{Shakya:2015xnx}, or hadrons~\cite{Bhattiprolu:2022sdd} annihilate to produce the dark matter. The same is possible for photons. The low reheat temperature region of parameter space can be probed in a model-dependent way by the Large Hadron Collider (LHC), via production of the new heavy states. It can also be probed by the next-generation of direct detection experiments, such as XLZD~\cite{XLZD:2024nsu}. We also derive bounds from Supernova 1987A. A similar freeze-in scenario was also recently considered in Refs.~\cite{Barman:2024nhr,Barman:2024tjt}, which investigated the potential sensitivity of a future lepton collider to a model in which dark matter couples equally to all electroweak gauge bosons.

The structure of this paper is as follows. In Sec.~\ref{sec: photonic freeze-in}, we describe dark matter production via photonic freeze-in, considering both instantaneous and non-instantaneous reheating scenarios. In Sec.~\ref{sec: DM couplings} we derive the radiatively generated couplings to SM fermions. Constraints from direct detection, astrophysics, cosmology, and colliders are then discussed in Sec.~\ref{sec: limits}.

\section{Photonic freeze-in}
\label{sec: photonic freeze-in}

\subsection{Scalar dark matter}

We first focus on photonic freeze-in of real scalar dark matter. At low energies, the dark matter $\chi$ couples to the SM via the effective interaction
\begin{equation}
\label{eq: DM-photon interaction scalar}
    \mathcal{L}_{\chi\gamma}^{s}=\frac{e^2}{16\pi^2}\frac{1}{\Lambda_s^2}\frac{1}{2}\chi^2 F_{\mu\nu}F^{\mu\nu} \,,
\end{equation}
where $F_{\mu\nu}$ is the electromagnetic field strength tensor. The factor of $e^2/16\pi^2$ is motivated by the fact that this interaction can be generated radiatively in the UV via heavy, electromagnetically charged particles, the mass of which is parametrically related to the scale $\Lambda_s$. We discuss possible UV completions in \cref{sec: UV completion}.

For sufficiently large values of $\Lambda_s$, the dark matter remains out of equilibrium with the SM thermal bath in the early universe. The interaction in \cref{eq: DM-photon interaction scalar} then sources a freeze-in population of dark matter, whose number density, $n_{\chi}$, is obtained by solving the Boltzmann equation
\begin{equation}
    \label{eq: General BE}
    \dot{n}_{\chi}+3Hn_{\chi}=2\sum\langle\sigma v\rangle_{i+j\rightarrow \chi + \chi + ...}n_i n_j \,,
\end{equation}
where $\langle \sigma v\rangle_{i+j\rightarrow \chi +\chi + ...}$ is the thermally-averaged cross-section times velocity for the  dark matter production process $i + j\rightarrow\chi + \chi + ...$, $n_i$ is the number density of species $i$, and the sum is over all relevant $\chi$-production processes. Similarly to the case of axion-like particle (ALP) freeze-in~\cite{Jain:2024dtw}, the interaction in \cref{eq: DM-photon interaction scalar} sources three main production processes: 
\begin{enumerate}
    \item photon annihilation ($\gamma\gamma\rightarrow\chi\chi$), 
    \item Primakoff-like production ($\gamma  X\rightarrow \chi  \chi+X$),
    \item particle-antiparticle annihilation ($\bar{X}X\rightarrow \chi\chi\gamma$),
\end{enumerate}
where $X$ represents any electromagnetically charged particles present in the thermal bath.

Relative to (i), processes (ii) and (iii) are suppressed by an additional phase space factor and power of $\alpha$. Consequently, photon annihilation dominates over the other channels. Furthermore, since all three processes have the same temperature dependence away from particle mass thresholds, this conclusion is independent of the reheat temperature, $T_R$. We therefore neglect processes (ii) \& (iii) for the remainder of this work.

\begin{figure*}
    \centering
    \includegraphics[width=0.49\linewidth]{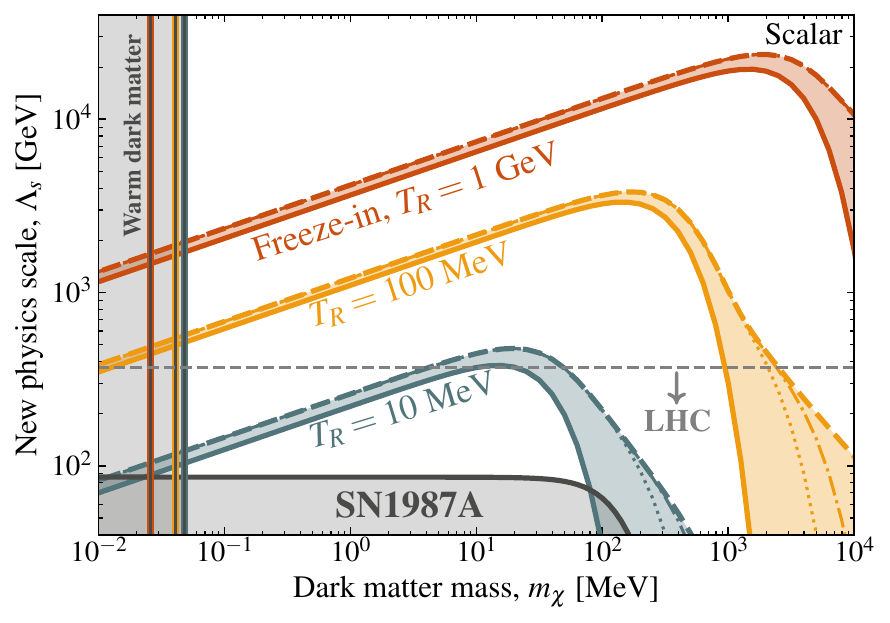}
    \includegraphics[width=0.49\linewidth]{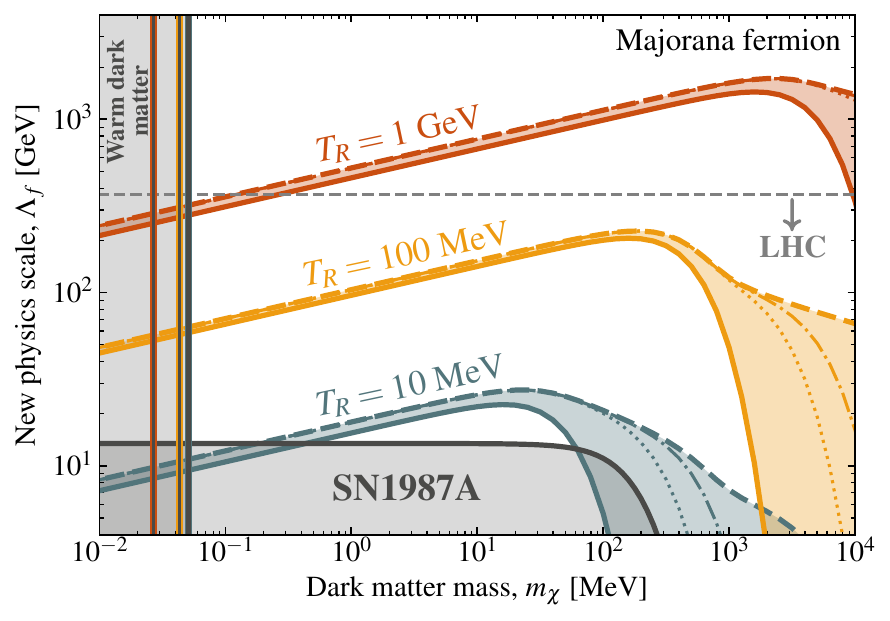}
    \caption{(Left) Values of $\Lambda_s$ required to saturate the dark matter relic abundance as functions of the dark matter mass, $m_{\chi}$, for $T_R=10$\,MeV (\textcolor{GreyBlue}{\textbf{blue}}), $100$\,MeV (\textcolor{Gold}{\textbf{yellow}}) and $1$\,GeV (\textcolor{BurntOrange}{\textbf{orange}}). The solid lines assume instantaneous reheating, while the dotted, dot-dashed and dashed lines correspond to non-instantaneous reheating with $T_{\rm{max}}/T_R=5$, $10$ and $100$. Constraints from SN1987A and limits on warm dark matter are denoted by \textcolor{WDMgrey}{\textbf{grey}} regions, with the latter also depending on the reheat temperature. The dashed line indicates the region that could be probed with LHC searches; as discussed in the main body, this is included for illustrative purposes and depends on the UV completion. (Right) The same, but for Majorana fermion dark matter with a different y-axis scale.}
\label{fig: FI results}
\end{figure*}

Using the standard expression for the thermally averaged-cross-section~\cite{Gondolo:1990dk}, we have that
\begin{equation}
\begin{split}
    \label{eq: thermally averaged CS 1}
    \langle\sigma v\rangle_{\gamma\gamma\rightarrow\chi\chi}n_{\gamma}^2=\frac{T}{512\pi^5 (2!)^2}&\int_{4m_{\chi}^2}^{\infty}ds\, \sqrt{s}K_1(\sqrt{s}/T)\\
    &\sqrt{1-\frac{4m_{\chi}^2}{s}}\sum_{\rm{pols}}|\mathcal{M}|^2 \,,
\end{split}
\end{equation}
where the integral is over the Mandelstam variable $s$. For simplicity, we have assumed that the photons follow a Maxwell-Boltzmann distribution and we have included symmetry factors for identical inital and final-state particles. In the above expression, $T$ is the temperature of the SM bath, $K_1$ is a modified Bessel function of the second kind and $\sum_{\rm{pols.}}|\mathcal{M}|^2$ is the squared-matrix element for photon annihilation to dark matter summed over all photon polarisations. Substituting in
\begin{equation}
    \label{eq: scalar annihilation Me}
    \sum_{\rm{pols}}|\mathcal{M}|^2=\frac{\alpha^2}{2\pi^2 \Lambda_s^4}s^2 \,,
\end{equation}
and introducing the integration variable $x=s/T^2$, we may re-write \cref{eq: thermally averaged CS 1} as
\begin{equation}
    \label{eq: thermally averaged cross-section scalar}
     \langle\sigma v\rangle_{\gamma\gamma\rightarrow\chi\chi}n_{\gamma}^2=\frac{\alpha^2 T^8}{64\pi^5\Lambda_s^4}\mathcal{I}_s(T, m_{\chi}) \,,
\end{equation}
with
\begin{equation}
\begin{split}
    \mathcal{I}_s(T, m_{\chi})\equiv\frac{1}{64\pi^2}&\int_{4m_{\chi}^2/T^2}^{\infty}dx \,x^{5/2}K_1(\sqrt{x})\\
    &\times\sqrt{1-\frac{4m_{\chi}^2}{xT^2}} \,,
\end{split}
\end{equation}
which asymptotes to approximately $1.22$ as $m_{\chi}/T \to 0$.

\subsubsection{Instantaneous reheating}
\label{sec: instantaneous reheating}

The fact that the thermally averaged cross-section in \cref{eq: thermally averaged cross-section scalar} is proportional to $T^8$ guarantees that the freeze-in production is UV-dominated, i.e.\ the final value of $n_{\chi}$ depends sensitively on the reheat temperature of the universe, $T_R$, and, potentially, the dynamics of reheating. For now, we assume instantaneous reheating, with $n_{\chi}(T_R)=0$, but we shall relax this assumption in the next section.

The Boltzmann equation \eqref{eq: General BE} is most easily solved by re-writing it in terms of the dark matter yield, $Y_{\chi}\equiv n_{\chi}/s$, where $s$ is the entropy density of the universe. Assuming the universe evolves adiabatically, this yields
\begin{equation}
    \label{eq: BE yield}
    \frac{dY_{\chi}}{dT}=-\frac{2\langle \sigma v\rangle_{\gamma\gamma\rightarrow\chi\chi} n_{\gamma}^2}{HsT}\bigg(1+\frac{T}{3}\frac{d\ln g_{*s}}{dT}\bigg) \,,
\end{equation}
where $g_{*s}(T)$ is the effective number of relativistic entropic degrees of freedom. We adopt values of $g_{*s}$ and $g_{*}$ from Ref.~\cite{Saikawa:2020swg}. The yield at temperature $T$ is therefore given by
\begin{equation}
    \label{eq: chi yield}
    Y_{\chi}(T)=\frac{\alpha^2}{32\pi^5\Lambda_s^4}\int_{T}^{T_R}\frac{d\tau\, \tau^7 \mathcal{I}_s(\tau, m_{\chi})}{H(\tau) s(\tau)}\bigg(1+\frac{\tau}{3}\frac{d\ln g_{*s}}{d\tau}\bigg) \,.
\end{equation}
In order for $\chi$ to account for the entire relic abundance of dark matter, we require this to be equal to~\cite{Planck:2018vyg}
\begin{equation}
    \label{eq: DM yield}
    Y_\text{DM} = 4.37\times10^{-7}\bigg(\frac{1~\rm{MeV}}{m_{\chi}}\bigg)
\end{equation}
at late times.

In the left panel of Fig.~\ref{fig: FI results}, the solid coloured lines show the values of the new physics scale, $\Lambda_s$, required to saturate the above equality for several values of the reheat temperature: $T_R=10$\,MeV (\textcolor{GreyBlue}{\textbf{blue}}), $100$\,MeV (\textcolor{Gold}{\textbf{yellow}}) and $1$\,GeV (\textcolor{BurntOrange}{\textbf{orange}}). We focus on low reheat temperatures because although photonic freeze-in remains perfectly viable for larger $T_R$, the required values of $\Lambda_s$ become significantly larger and are therefore increasingly difficult to probe experimentally. We note, however, that baryogenesis favours regions of parameter space with higher $T_R$ (or, in the case of non-instantaneous reheating to be discussed in the next section, higher $T_\text{max}$). Big Bang nucleosynthesis provides a strict lower bound on $T_R$, since it is no longer consistent with the observed light element abundances for reheat temperatures below $\sim5$\,MeV~\cite{deSalas:2015glj,Hasegawa:2019jsa}.

We also show in Fig.~\ref{fig: FI results} the constraint from dark matter production in SN1987A and the warm dark matter bound from structure formation. The latter depends on the reheat temperature, as shown by the vertical coloured lines. The dashed grey line indicates the region of parameter space that could be probed by LHC searches for the heavy, electromagnetically charged particles that are responsible for generating the operator in \cref{eq: DM-photon interaction scalar} in the UV. These are all discussed in detail in Sec.~\ref{sec: limits}. 

\subsubsection{Non-instantaneous reheating}
\label{sec: non-instantaneous reheating}

In realistic models of inflation, reheating occurs over some finite period rather than instantaneously. In such scenarios, the nascent SM bath reaches a maximum temperature, $T_{\rm{max}}$, shortly after the end of inflation, which can, depending on the inflationary model, be orders of magnitude larger than $T_R$~\cite{Allahverdi:2010xz}. Although access to a higher maximum temperature is seemingly important for a UV-dominated production process like photonic freeze-in, the continued damping of the inflaton field during reheating increases the entropy of the SM bath, diluting any population of dark matter produced at higher temperatures. Whether or not the final dark matter yield has a significant dependence on $T_{\rm{max}}$ depends on how these two effects compete with one another.

This phenomenon has been investigated thoroughly in Refs.~\cite{Garcia:2017tuj, Garcia:2020eof, Bhattiprolu:2022sdd}. In the absence of any direct interaction between the inflaton and dark matter, and assuming the inflaton energy-density redshifts like matter during reheating (as is the case when its potential is quadratic about its minimum), the contribution to the total dark matter yield from temperatures above $T_R$ is given by~\cite{Garcia:2017tuj, Bhattiprolu:2022sdd}
\begin{equation}
    \label{eq: yield NI}
    \begin{split}
    Y_{\chi}^{\rm{N.I.}}\simeq\frac{216\sqrt{10} M_{\rm{pl}}}{\pi^3}&\frac{g_{*}^{5/2}(T_R)}{g_{*s}(T_R)}T_R^7\int_{T_R}^{T_{\rm{max}}} \frac{dT \langle \sigma v\rangle n_{\gamma}^2}{T^{13} g_{*}^3(T)}\\
    &\times
    \begin{cases}
        0.8\quad \rm{(scalar\ DM)}\\
        0.45\quad \rm{(fermionic\ DM)}.
    \end{cases}
    \end{split}
\end{equation}
The total dark matter yield is then obtained by adding this contribution to that obtained from \cref{eq: chi yield}.

Defining $n$ as the power of temperature in the thermally-averaged cross-section times velocity, i.e. $\langle\sigma v\rangle n_{\gamma}^2\propto T^{n+6}$, it is clear from \cref{{eq: yield NI}} that the dark matter yield is determined by $T_R$ rather than $T_{\rm{max}}$ when $n<6$~\cite{Garcia:2017tuj}. If $T_R > m_\chi$, we have $n=2$ for the photonic freeze-in of scalar dark matter and the above integral will be dominated by temperatures close to $T_R$. On the other hand, when $T_R < m_{\chi}$, the thermally-averaged cross-section is Boltzmann suppressed at $T_R$ and the details of reheating can significantly affect the dark matter yield. 

The shaded bands in Fig.~\ref{fig: FI results} show the ranges of $\Lambda_s$ that yield the correct dark matter abundance for $T_R=10$\,MeV (\textcolor{GreyBlue}{\textbf{blue}}), $100$\,MeV (\textcolor{Gold}{\textbf{yellow}}) and $1$\,GeV (\textcolor{BurntOrange}{\textbf{orange}}) with non-instantaneous reheating. The dotted, dot-dashed, and dashed lines correspond to $T_{\rm{max}}/T_R=5$, $10$ and $100$, respectively.

For low dark matter masses ($m_\chi<T_R$), the results for non-instantaneous reheating are within an order one factor of their instantaneous counterparts. This is because the additional dark matter production in the non-instantaneous case is concentrated at temperatures just above $T_R$ and is essentially the same for all values of $T_{\rm{max}}/T_R$. In contrast, when $m_\chi>T_R$, dark matter production during reheating peaks around the lowest temperature for which it is no longer Boltzmann-suppressed, i.e. at $T\sim m_{\chi}$, and can significantly enhance the dark matter yield. For this reason, the $T_{\rm{max}}/T_R=5, 10$ curves closely follow the $T_{\rm{max}}/T_R=100$ result until $m_{\chi}>T_{\rm{max}}$ at which point they fall off exponentially. Finally, note that already for $T_{\rm{max}}/T_R=100$ we have that $T_{\rm{max}}>m_\chi$ across the entire dark matter mass range we consider for both $T_R=100$\,MeV and $T_R=1$\,GeV ; hence, increasing the ratio $T_{\rm{max}}/T_R$ further would have a negligible impact on the dark matter abundance and the required value of $\Lambda_s$.

\subsection{Fermionic dark matter}

Photonic freeze-in can be equally applied to fermionic dark matter. In this scenario, the relevant low-energy interaction is
\begin{equation}
    \label{eq: DM-photon interaction fermion}
    \mathcal{L}_{\chi\gamma}^{f}=\frac{e^2}{16\pi^2}\frac{1}{2\Lambda_f^3}\bar{\chi}\chi F_{\mu\nu}F^{\mu\nu} \,.
\end{equation}
We take the dark matter to be a Majorana fermion. The main production process is again $\gamma\gamma\rightarrow \chi\chi$, with total squared matrix element
\begin{equation}
\label{eq: fermion square matrix element}
    \sum_{\rm{spin}}|\mathcal{M}|^2 = \frac{\alpha^2}{\pi^2\Lambda_f^6}\, s^2\big(s-4m_{\chi}^2\big) \,.
\end{equation}
This corresponds to a thermally averaged cross-section
\begin{equation}
\label{eq: thermally averaged cross-section fermion}
    \langle\sigma v \rangle_{\gamma\gamma\rightarrow\chi\chi} n_{\gamma}^2=\frac{\alpha^2 T^{10}}{32\pi^5 \Lambda_f^6} \mathcal{I}_f(T, m_{\chi}) \,,
\end{equation}
where now
\begin{equation}
    \mathcal{I}_f(T, m_{\chi})\equiv \frac{1}{64\pi^2}\int_{4m_{\chi}^2/T^2}^{\infty} dx K_1(\sqrt{x})x^2 \big(x-4m_{\chi}^2/T^2\big)^{3/2} \,.
\end{equation}

Using the above expressions, we have calculated the values of $\Lambda_f$ needed to saturate the dark matter relic abundance for both instantaneous and non-instantaneous reheating. The results are shown in the right panel of Fig.~\ref{fig: FI results}. As previously, the solid curves correspond to instantaneous reheating with $T_R=10$\,MeV (\textcolor{GreyBlue}{\textbf{blue}}), $100$\,MeV (\textcolor{Gold}{\textbf{yellow}}) and $1$\,GeV (\textcolor{BurntOrange}{\textbf{orange}}), with non-instantaneous reheating populating the shaded regions. The dotted, dot-dashed and dashed lines correspond to $T_{\rm{max}}/T_R=5$, $10$ and $100$, respectively. 

The results are qualitatively similar to the case of scalar dark matter, discussed in the previous section. However, for fermionic dark matter, $\langle\sigma v \rangle$ is suppressed by an additional factor of $(T/\Lambda_f)^2$ and hence the new physics scales needed to achieve the correct dark matter abundance are correspondingly lower than their counterparts for scalar dark matter. Since $\langle\sigma v \rangle n_{\gamma}^2\propto T^{10}$ (i.e. $n=4$), the abundance of dark matter produced during non-instantaneous reheating is again independent of $T_{\rm{max}}$ when $m_{\chi} < T_{\rm{max}}$. As in the scalar case, deviations from the case of instantaneous reheating are therefore small for $m_{\chi}\lesssim T_R$, while the viable parameter space opens up significantly when $m_{\chi} > T_R$.

\section{Dark matter couplings to visible matter}
\label{sec: DM couplings}

\begin{figure}
    \centering
    \includegraphics[width=0.6\linewidth]{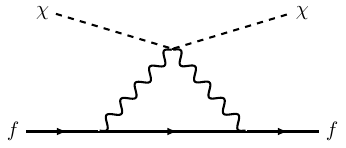}
    \caption{Feynman diagram that generates the coupling between the dark matter $\chi$ and the SM fermions $f$.}
    \label{fig: FD}
\end{figure}

So far, we have only considered the effective dark matter interaction with photons, relevant for photonic freeze-in. However, this interaction radiatively generates couplings to quarks and leptons via the UV-divergent diagram shown in Fig.~\ref{fig: FD}. The low-energy effective Lagrangian therefore also contains terms of the form
\begin{equation}
\label{eq: L_chif}
    \mathcal{L}_{\chi f} = g_{\chi f} \bar{f} f  \begin{cases}
        \frac{1}{2}\chi^2\quad \rm{(scalar)} \,, \\
        \frac{1}{2}\bar{\chi}\chi\quad \rm{(fermionic)} \,.
    \end{cases}
\end{equation}
In the following, we assume that the coupling $g_{\chi f}$ vanishes at the UV scale $\Lambda_{s,f}$, motivated by minimal UV completions in which this interaction is only generated at the two-loop level. Evolving the coupling to lower scales using the one-loop renormalisation group equation for $g_{\chi f}$ yields
\begin{equation}
\label{eq: DM-fermion coupling}
    g_{\chi f}(\mu)=-\frac{3 Q_f^2\alpha^2 m_f}{4\pi^2}
    \begin{cases}
    \ln\big(\frac{\Lambda_{s}^2}{\mu^2}\big)\frac{1}{\Lambda_s^2}\quad \rm{(scalar)} \,, \\
       \ln\big(\frac{\Lambda_{f}^2}{\mu^2}\big)\frac{1}{\Lambda_f^3}\quad \rm{(fermionic)} \,,
    \end{cases}
\end{equation}
where $Q_f$ is the electromagnetic charge of the relevant SM fermion and $\mu$ is the renormalisation scale. Note that the coupling $g_{\chi f}$ is proportional to the fermion mass. This means the dark matter coupling to electrons is suppressed by $m_e$, and so is less phenomenologically relevant than the hadronic couplings.

We will be primarily interested in the low-energy coupling of the dark matter to nucleons, which we derive as follows.  First, integrating out the heavy quarks at their corresponding mass scales yields the dark-matter--gluon interaction
\begin{equation}
\begin{split}
    \mathcal{L}_{\chi g} = \frac{\alpha_s}{8\pi}\bigg(\sum_{q=c,b,t}& \frac{-2 g_{\chi q}(m_q)}{3m_q}\bigg)G_{\mu\nu}^a G^{a\, \mu\nu}\\
    &\times\begin{cases}
        \frac{1}{2}\chi^2\quad \rm{(scalar)} \,, \\
        \frac{1}{2}\bar{\chi}\chi\quad \rm{(fermionic)} \,,
    \end{cases}
\end{split}
\end{equation}
where $\alpha_s$ and $G_{\mu\nu}^a$ are the strong coupling constant and field-strength tensor respectively, and $g_{\chi q}$ is evaluated at $\mu=m_q$. Taking this together with the dark matter coupling to the light quarks and running\footnote{Note that the operator $\alpha_s G G$ is approximately RGE invariant~\cite{Grinstein:1988wz}.} down to $\mu\sim1\,$GeV, we can then match onto the dark-matter--nucleon couplings in Heavy Baryon Chiral Perturbation Theory (HBChPT)~\cite{JENKINS1991558}. Following Ref.~\cite{Bishara:2016hek}, the resulting dark-matter--nucleon coupling is
\begin{equation}
    \label{eq: DM-nucleon coupling}
    g_{\chi N}=\frac{1}{9}m_G\bigg(\sum_{q=c,b,t} \frac{-2 g_{\chi q}(m_q)}{3m_q}\bigg) +\sum_{q=u,d,s}g_{\chi q}(\Lambda_{\rm{QCD}}) b_0 \,,
\end{equation}
where the interaction takes the same form as in \cref{eq: L_chif}. The low-energy constants $b_0=-3.2\pm 0.3$ and $m_G=(847\pm 8)$\,MeV were evaluated in Ref.~\cite{Cox:2024rew} using lattice data~\cite{FlavourLatticeAveragingGroupFLAG:2021npn}. This leads to the low-energy dark-matter--nucleon scattering cross-sections
\begin{equation}
    \label{eq: DM-nucleon cross-section}
    \sigma_{\chi N} = \frac{g_{\chi N}^2}{4\pi}\frac{\mu_{\chi N}^2}{m_{\chi}^2}\begin{cases}
        1\quad \rm{(scalar)} \,,\\
        4m_{\chi}^2 \quad \rm{(fermionic)} \,,
    \end{cases}
\end{equation}
where $\mu_{\chi N}$ is the dark-matter--nucleon reduced mass.

\section{Phenomenology and Constraints}
\label{sec: limits}

There is a suite of constraints from direct detection, astrophysics, cosmology, and colliders that have the potential to constrain the photonic freeze-in scenario. Some of these directly probe the interactions with photons in \cref{eq: DM-photon interaction scalar,eq: DM-photon interaction fermion} responsible for freeze-in, while others are sensitive to the loop-induced coupling to fermions in \cref{eq: DM-fermion coupling}. It turns out that in much of the parameter space, high-energy collider searches for the new charged states that generate the dark matter coupling to photons in the UV are potentially the most powerful way to probe this scenario.

\subsection{UV Completions \& Collider Searches}
\label{sec: UV completion}

In this section, we briefly comment on possible UV completions of the photonic freeze-in scenario. These generically\footnote{The operators in \cref{eq: DM-photon interaction scalar,eq: DM-photon interaction fermion} can also be generated by UV completions without new charged states, for example via the tree-level exchange of a neutral scalar that mixes with the Higgs. However, such scenarios also generate unsuppressed couplings to quarks and/or leptons and therefore lie outside of the photonic freeze-in paradigm we are considering.} require new heavy, electromagnetically charged degrees of freedom that, when integrated out, generate the effective operators in Eqs.~\eqref{eq: DM-photon interaction scalar} or \eqref{eq: DM-photon interaction fermion}.

\begin{figure}
    \centering
    \includegraphics[width=0.2\textwidth]{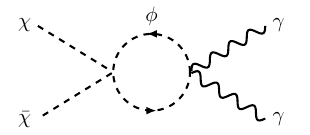}
    \hspace{1.5em} 
    \includegraphics[width=0.2\textwidth]{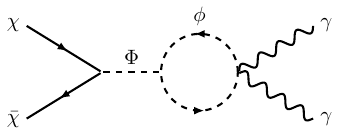} \\
    \includegraphics[width=0.2\textwidth]{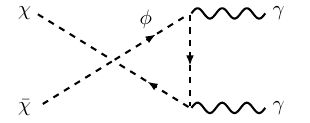}
    \hspace{1.5em}
    \includegraphics[width=0.2\textwidth]{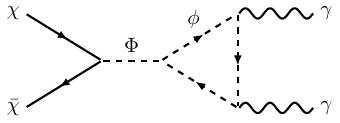}
    \caption{Example minimal UV completions for scalar (left) and fermionic (right) dark matter, where $\phi$ and $\Phi$ are heavy charged and neutral scalars, respectively. (Additional diagrams with exchanged photons are not shown.)}
    \label{fig: UV completions}
\end{figure}

The minimal UV completions for both scalar and fermionic dark matter generate the effective dark-matter--photon interaction at one loop. In the case of scalar dark matter, this requires a single new charged state, which could be either a complex scalar or a Dirac fermion. For fermionic dark matter, the minimal UV completion requires two additional heavy scalars, one of which is charged. Example diagrams are shown in \cref{fig: UV completions}.

UV completions will generate additional operators in the low energy EFT, beyond those in \cref{eq: DM-photon interaction scalar,eq: DM-photon interaction fermion}, which could potentially modify the low-energy dark matter phenomenology. Of particular note are the lower-dimension operators that couple the DM to a single photon field (see e.g.~\cite{Kavanagh:2018xeh}). There are no such operators for real scalar dark matter, but for Majorana fermion dark matter there is the dim-6 anapole operator. (If one considers non-self-conjugate dark matter there are also the charge radius and dipole operators). For the UV completions we have presented in \cref{fig: UV completions}, it is straightforward to show that interactions with a single photon only arise at two-loops and can hence be suppressed. 

If the exotic(s) were to transform non-trivially under $SU(3)$ colour, then integrating them out would also generate an effective coupling of the dark matter to gluons. This would enable additional, hadronic dark matter production channels, such as $\pi^+\pi^-\to\chi\chi$, that may dominate over the photonic channel. The low-temperature freeze-in of hadrophilic dark matter was recently considered in Ref.~\cite{Bhattiprolu:2022sdd}.

Collider searches, in particular at the LHC, are potentially sensitive to these new charged states, which can be produced through SM gauge interactions. Here, we focus on the minimal models in which these new states are charged only under $U(1)_Y$; states that have non-trivial $SU(2)_L$ or $SU(3)$ quantum numbers will generally be subject to significantly stronger constraints due to their larger production cross-sections. 

In general, the collider signatures will depend on the details of the UV completion that determine how the exotic state(s) decay. However, there is a broad class of UV completions with the same signature; these are models in which the exotics have charges that are not integer multiples of $1/3$ and are therefore stable. There are dedicated LHC searches for stable charged particles. In particular, the CMS search in Ref.~\cite{CMS:2024eyx} places bounds that reach up to masses of 640\,GeV (370\,GeV~\cite{Koren:2024xof}) for fermions (scalars) with charges $\approx 2/3$; however, the constraints are significantly weaker for states with smaller or larger charges. A more general overview of the collider bounds on fractionally-charged particles, including bounds on non-trivial $SU(2)_L$ representations, can be found in Ref.~\cite{Koren:2024xof}.

We have indicated the potential of collider searches to constrain the photonic freeze-in parameter space with the dashed grey line in Fig.~\ref{fig: FI results}, which corresponds to $\Lambda_{s,f}=370$\,GeV, motivated by the CMS search discussed above. While this is purely indicative, with the actual bounds sensitive to the specific UV completion, it nevertheless highlights the fact that collider searches have the potential to strongly constrain photonic freeze-in at low reheat temperatures.

\subsection{Direct Detection}

\begin{figure*}
    \centering
    \includegraphics[width=0.49\linewidth]{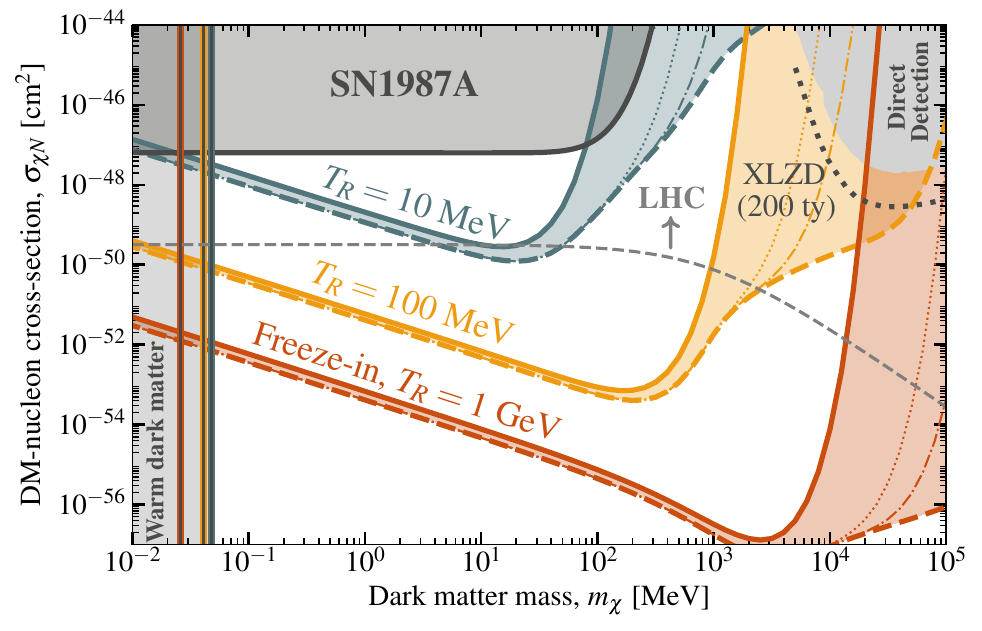}
    \includegraphics[width=0.49\linewidth]{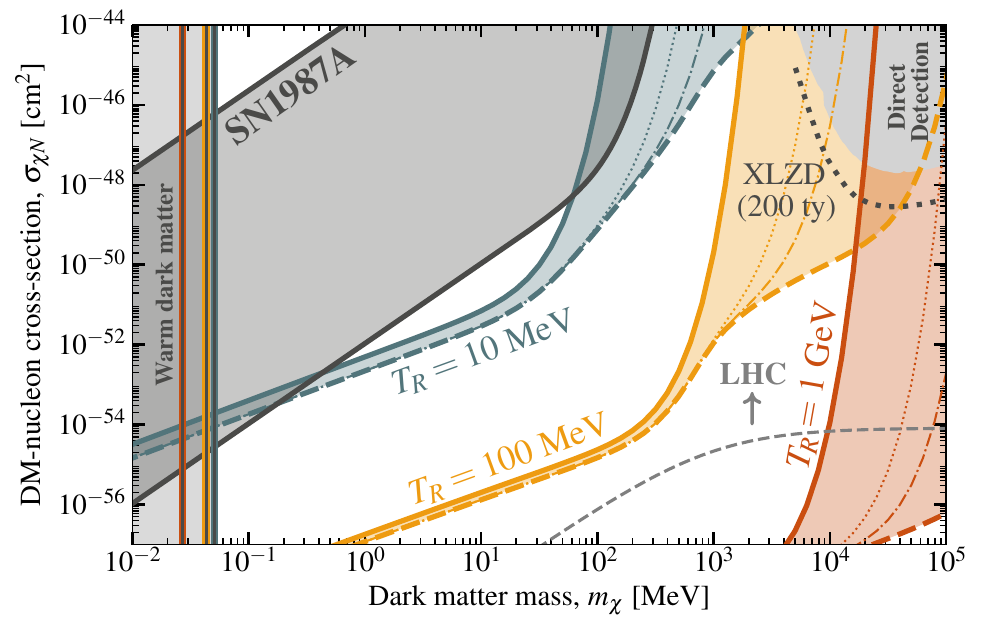}
    \caption{(Left) Scalar dark-matter--proton cross-sections for $T_R=10$\,MeV (\textcolor{GreyBlue}{\textbf{blue}}), $100$\,MeV (\textcolor{Gold}{\textbf{yellow}}) and $1$\,GeV (\textcolor{BurntOrange}{\textbf{orange}}). Thick solid and dashed lines assume instantaneous reheating and non-instantaneous reheating with $T_{\rm{max}}/T_R=100$, respectively, while $T_{\rm{max}}/T_R=5$ and $10$ are included as thin dotted and dot-dashed lines. Shaded grey regions represent bounds from direct detection, SN1987A and warm dark matter limits (which depend on $T_R$). The dashed grey line indicates the potential sensitivity of LHC searches, which, however, depends on the UV completion. The dotted grey line shows the projected future sensitivity of direct detection with XLZD. (Right) The same as the left panel, but for Majorana fermion dark matter.}
    \label{fig: proton direct detection}
\end{figure*}

Direct detection experiments can also probe the photonic freeze-in scenario via the radiatively generated couplings to fermions in \cref{eq: DM-fermion coupling}. The linear-dependence of $g_{\chi f}$ on $m_f$ implies that photonic freeze-in dark matter is best searched for through its couplings to nucleons, rather than electrons. 

In the left (right) panel of Fig.~\ref{fig: proton direct detection} we show the dark-matter--nucleon cross-sections corresponding to the values of $\Lambda_s$ ($\Lambda_f$) required for photonic freeze-in. Once again, results for $T_R=10$\,MeV, $100$\,MeV and $1$\,GeV are shown in \textcolor{GreyBlue}{\textbf{blue}}, \textcolor{Gold}{\textbf{yellow}}, and \textcolor{BurntOrange}{\textbf{orange}}, respectively, alongside limits from warm dark matter and SN1987A, which are discussed in Secs.~\ref{sec: cosmology} \&~\ref{sec: stars}. Predictions for instantaneous reheating are given by solid lines, while those for non-instantaneous reheating with $T_{\rm{max}}/T_R=5$, $10$ and $100$ are indicated by dotted, dot-dashed and dashed lines, respectively. Existing experimental limits on the spin-independent dark-matter--proton cross-section from LZ~\cite{LZCollaboration:2024lux}, PandaX~\cite{PandaX:2024qfu}, XENONnT~\cite{XENON:2024hup} and DarkSide~\cite{DarkSide:2022dhx} are included in \textcolor{DDgrey}{\textbf{grey}}, while the 200~ty projection from XLZD~\cite{XLZD:2024nsu} is indicated by a dotted grey line.

It's clear from Fig.~\ref{fig: proton direct detection} that current direct detection experiments are only sensitive to the region where $m_{\chi}>T_R$ for both scalar and fermionic dark matter. In this region dark matter production at $T_R$ is Boltzmann-suppressed and hence $\sigma_{\chi N}$ depends sensitively on the nature of reheating.

The region where $m_{\chi}<T_R$ is, however, very challenging to probe with direct detection. Only the lowest temperature reheating scenarios are plausibly within the reach of future experiments. While a number of experimental strategies have been suggested to search for dark matter in this low mass regime, mostly exploiting dark-matter-induced phonon excitations in polar materials or superfluid helium~\cite{Knapen:2016cue, Knapen:2017ekk, Campbell-Deem:2019hdx, Griffin:2019mvc, Griffin:2020lgd, Campbell-Deem:2022fqm, Baker:2023kwz, Ashour:2024xfp}, optimistic estimates place their sensitivities at around $\sigma_{\chi p}\sim 10^{-44}$--$10^{-45}$~cm$^{2}$, several orders of magnitude larger than our predictions for scalar dark matter with even the lowest possible reheat temperatures.

\subsection{Cosmology}
\label{sec: cosmology}

The most important cosmological constraint on this scenario comes from structure formation.
The very feeble interaction strengths inherent to freeze-in imply that other typical cosmological constraints on dark matter, including those from dark matter annihilation leading to energy deposition into the SM bath during different cosmological epochs, and the presence of additional relativistic species during Big Bang nucleosynthesis are not relevant.

If dark matter produced via photonic freeze-in is too light, its free-streaming length will be sufficiently large to prevent dark matter clustering on small scales. This manifests as a cutoff in the matter power spectrum at small Fourier modes, which can be constrained using, e.g., measurements of the Ly-$\alpha$ forest~\cite{Ballesteros:2020adh, Villasenor:2022aiy} or the abundance of Milky Way satellites~\cite{DES:2020fxi}.

These bounds are typically reported as lower limits on the masses of thermal dark matter particles, $m_{\rm{WDM}}$, which can be translated to arbitrary non-thermal dark matter phase space distributions. Specifically, Ref.~\cite{Ballesteros:2020adh} finds that limits on scalar and fermionic dark matter for processes like photonic freeze-in are given by
\begin{equation}
\begin{split}
    m_{\chi}\gtrsim& \bigg(\frac{m_{\rm{WDM}}}{3\ \rm{keV}}\bigg)^{4/3}\bigg(\frac{106.75}{g_{*s}^{\rm{R}}}\bigg)^{1/3}\\
    &\times
    \begin{cases}
    8.01\ \rm{keV}, \quad \rm{bosons}\\
    8.40\ \rm{keV}, \quad \rm{fermions}\\ 
    \end{cases}
\end{split}
\end{equation}
where $g_{*s}^{\rm{R}}$ is the number of relativistic entropic degrees of freedom at the time of reheating. Taking the most recent Dark Energy Survey limit on warm dark matter, $m_{\rm{WDM}}>6.5$\,keV~\cite{DES:2020fxi}, we obtain bounds of $m_{\chi}\gtrsim 42$\,keV (44\,keV) for scalar (fermionic) dark matter with a reheat temperature of $T_R=100$\,MeV. The limits for $T_R=10,\,100$ and $100$\,MeV are included in Figs.~\ref{fig: FI results}~and~\ref{fig: proton direct detection}.

\subsection{Astrophysics}
\label{sec: stars}

The leading astrophysical constraints on the operators in \cref{eq: DM-photon interaction scalar,eq: DM-photon interaction fermion} arise from stellar observations, such as the populations of horizontal branch stars and the neutrino signal associated with SN1987A. These give bounds on the energy loss via the production and escape of dark matter from the relevant stellar region, which is quantified via the rate of energy loss to dark matter per unit mass, $\epsilon_{\chi}=\sum_i \epsilon_{\chi}^i$, where the sum is over all contributing production processes. For horizontal branch stars, we impose the bound $\epsilon_{\chi}\lesssim 10$~erg g$^{-1}$ s$^{-1}$ for $T=10^8$~K $\simeq 8.6$\,keV and $\rho=10^{4}$~g cm$^{-3}$~\cite{Raffelt:1996wa}. The neutrino signal detected following SN1987A provides a similar, approximate constraint of $\epsilon_{\chi}\lesssim 10^{19}$~erg g$^{-1}$ s$^{-1}$, assuming average progenitor core conditions of $T=30$\,MeV and $\rho=3\times10^{14}$~g cm$^{-3}$~\cite{Raffelt:1996wa}. 

The most important process contributing to energy loss is $\gamma\gamma\to \chi\chi$. We also considered dark matter production via the \textit{Primakoff-like} process ($\gamma X \rightarrow \chi \chi X$) and nuclear bremsstrahlung ($N N \rightarrow N N \chi \chi$), but found both were over an order of magnitude less efficient than photon annihilation for the same $\Lambda_{s,f}$ (see App.~\ref{app: stellar energy-loss}). 

For the scalar dark matter operator in \cref{eq: DM-photon interaction scalar}, the energy-loss rate per unit mass via photon annihilation is given by
\begin{equation}
    \label{eq: photon annihilation energy-loss scalar}
    \epsilon^{\rm{ann},s}_{\chi}=\frac{\alpha^2}{64\pi^5 \Lambda_s^4}\frac{T^9}{\rho}F_{s}(m_{\chi}/T) \,,
\end{equation}
where $T$ and $\rho$ are the local stellar temperature and density. The function $F_s(m_{\chi}/T)$, defined in App.~\ref{app: stellar energy-loss}, is $\mathcal{O}(1)$ when $m_{\chi}\lesssim T$ and exponentially decays for $m_{\chi}\gg T$. For dark matter masses lower than the relevant core temperature, horizontal branch stars then exclude $\Lambda_s \lesssim 12.1$\,GeV, while SN1987A excludes $\Lambda_s\lesssim 86.1$\,GeV. The latter limit is shown in grey in Figs.~\ref{fig: FI results} and ~\ref{fig: proton direct detection}. These limits do not extend to arbitrarily low $\Lambda_{s}$, as the dark matter will become trapped and thermalise within the stellar core; we discuss this further below.

For Majorana fermion dark matter, we find that
\begin{equation}
    \label{eq: photon annihilation energy-loss fermion}
    \epsilon_{\chi}^{\rm{ann},f}=\frac{\alpha^2}{32\pi^5\Lambda_f^6}\frac{T^{11}}{\rho}F_{f}(m_{\chi}/T) \,,
\end{equation}
with the function $F_f(m_{\chi}/T)$ also defined in App.~\ref{app: stellar energy-loss}. Applying the relevant conditions in horizontal branch stars and the SN1987A proto-neutron star (PNS) excludes values of $\Lambda_f\lesssim 0.27$\,MeV and $\Lambda_f\lesssim 15.1$\,GeV, respectively. The SN1987A limit strongly constraints photonic freeze-in of fermionic dark matter for sub-MeV masses and $T_R=10\,$MeV (see Fig.~\ref{fig: FI results}).

\subsubsection{Trapping}
\label{sec: trapping}

If dark-matter--SM interactions are sufficiently strong, dark matter produced in stars may become trapped instead of escaping, which relaxes the energy-loss constraints. The SN1987A limit in this regime has been examined in Refs.~\cite{Chang:2018rso, DeRocco:2019jti}. Given sufficiently strong interactions, a thermal population of dark matter will form in the PNS core, which contributes a luminosity $L_{\chi}$ through the neutrino core radius $R_{\nu}\simeq 20$~km. Parameter space where $L_{\chi}$ is larger than the neutrino luminosity, $L_{\nu}=3\times10^{52}$~erg~s$^{-1}$, is ruled out~\cite{Raffelt:1996wa}. 

$L_{\chi}$ can be estimated by identifying the radius, $R_E$, and temperature at which dark matter thermally decouples from the PNS medium. As the interaction strength increases (i.e. decreasing $\Lambda_{s,f}$), this radius increases, with a correspondingly lower dark matter temperature and therefore $L_{\chi}$. 

Under the approximation of instantaneous decoupling, $R_E$ is obtained by solving~\cite{Shapiro-1983}
\begin{equation}
    \int_{R_E}^{\infty} dr' \big(\lambda_{\chi E}^{\rm{eff}}\big)^{-1} = \frac{2}{3},
\end{equation}
where the left-hand side is the optical depth for thermalisation and $\lambda_{\chi E}^{\rm{eff}}$ is the effective mean free path for energy-exchanging dark matter scattering. Because dark-matter nucleon scattering is inefficient at transferring energy ($T<m_N$), this should be identified with $\lambda_{\chi\gamma}^{\rm{eff}}$. However, Ref.~\cite{DeRocco:2019jti} found that the instantaneous decoupling approximation overestimates the dark matter luminosity compared with the results of a full Boltzmann transport equation. To ensure that our limit is conservative, we instead assume that dark matter remains thermal out to the larger nucleon scattering radius, obtained by replacing $\lambda_{\chi E}^{\rm{eff}}$ with $\lambda_{\chi N}=1/(n_N \sigma_{\chi N})$, where $n_N$ is the nucleon number density.

The dark matter luminosity is
\begin{equation}
    L_{\chi} = 4\pi R_E^2\frac{g_{\chi}}{8\pi^2}\int_0^{\infty}\frac{dp_{\chi}\, p_{\chi}^3}{e^{E_{\chi}/T_S}\pm 1},
\end{equation}
where $g_{\chi}$ is the number of degrees of freedom of the dark matter, $T_E$ is the temperature at $R_E$, and the $\pm$ in the denominator is appropriate for fermionic/scalar dark matter. We compute bounds using the Fischer $11M_{\odot}$, Fischer $18M_{\odot}$~\cite{Fischer:2016cyd} and Nakazato~\cite{Nakazato:2012qf} PNS profiles from~\cite{Chang:2018rso, DeRocco:2019jti} and use the most conservative limit ($18M_{\odot}$ profile). This defines the upper edge of the SN1987A excluded region in Fig.~\ref{fig: proton direct detection}.

\section{Conclusion}
\label{sec: conclusion}

We have introduced photonic freeze-in, a scenario where dark matter is produced via photon annihilation in the early universe. This is a very minimal possibility which, besides the dark matter itself, requires only an additional heavy, charged state in the UV. 

We have delineated the parameter space where this scenario can achieve the correct relic abundance as a function of the reheat temperature, including the impact of non-instantaneous reheating. The phenomenologically interesting regions of the parameter space correspond to low reheat temperatures. 

While we have focused on dark matter that couples to $F^{\mu\nu} F_{\mu\nu}$, photonic freeze-in could equally proceed via a coupling to $F^{\mu\nu} \tilde{F}_{\mu\nu}$. The phenomenology of such scenarios is very similar to those we have presented, with the exception that the direct detection cross-section becomes both spin-dependent and momentum-suppressed.

We have also outlined possible UV completions of the photonic freeze-in scenario for both scalar and fermionic dark matter. It would be interesting to undertake a more detailed model-building analysis in the future and to investigate the potential of LHC searches to test specific UV models. It would also be worthwhile to further study the dependence on the reheating dynamics.


\acknowledgments
We would like to thank Alexei Sopov for helpful comments. This work was supported in part by the Australian Research Council through the ARC Centre of Excellence for Dark Matter Particle Physics, CE200100008. P.C. is supported by the Australian Research Council Discovery Early Career Researcher Award DE210100446. 

\bibliographystyle{JHEP}
\bibliography{main}

\newpage
\appendix

\section{Stellar energy-loss rates}
\label{app: stellar energy-loss}

In this appendix, we derive the energy-loss rates used in Sec.~\ref{sec: limits} to calculate limits on the dark-matter--photon coupling from horizontal branch stars and SN1987A.

\subsection{Photon annihilation}

The rate of energy loss per unit mass due to dark matter production via photon annihilation ($\gamma (P_a) + \gamma (P_b) \rightarrow \chi (P_1) + \chi (P_2)$) is
\begin{equation}
\begin{split}
\epsilon_{\chi}^{\rm{ann}}=\frac{S}{2!}\frac{1}{\rho}\int& d\Pi_a d\Pi_b d\Pi_1 d\Pi_2 f_a f_b (E_1 + E_2) (2\pi)^4\\
    &\times \delta^4(P_a + P_b - P_1 - P_2)\sum_{\rm{d.o.f.}}|\mathcal{M}|^2\,,
\end{split}
\end{equation}
where $\rho$ is the local stellar density, $S$ is a symmetry factor equal to 1/2 (1) for real scalar or Majorana fermion (complex scalar or Dirac fermion) dark matter, and
\begin{equation}
    d\Pi_i=\frac{dp_i^3}{(2\pi)^3}\frac{1}{2E_i} \,.
\end{equation}
It has been assumed that the density of dark matter is negligible compared to the photons. Performing the phase space integral yields~\cref{eq: photon annihilation energy-loss scalar} and~\cref{eq: photon annihilation energy-loss fermion} with
\begin{equation}
    \begin{split}
        F_s(m_{\chi}/T)=\frac{1}{64\pi^2}&\int_{4m_{\chi}^2/T^2}^{\infty}dx x^{7/2}\sqrt{1-\frac{4m_{\chi}^2}{xT^2}}\\
        &\int_1^{\infty} dy y\sqrt{y^2-1}\exp(-\sqrt{x}y) \,,
    \end{split}
\end{equation}
and
\begin{equation}
    \begin{split}
        F_f(m_{\chi}/T)=\frac{1}{64\pi^2}&\int_{4m_{\chi}^2/T^2}^{\infty}dx x^{7/2}(x-4m_{\chi}^2)\sqrt{1-\frac{4m_{\chi}^2}{xT^2}}\\
        &\int_1^{\infty} dy y\sqrt{y^2-1}\exp(-\sqrt{x}y) \,.
    \end{split}
\end{equation}

\subsection{Primakoff-like process}

Next, we derive the energy-loss rate per unit mass via Primakoff-like dark matter production $\gamma (P_a) + X (P_b)\rightarrow \chi (P_1) + \chi (P_2) + X (P_3)$. For simplicity, we will calculate these under the assumption that $m_{\chi}\leq  T\ll m_{X}$, with $m_X$ the mass of the target particle. 

For a single target species $X$, this rate is given by
\begin{equation}
    \label{eq: Primakoff general energy loss}
    \begin{split}
        \epsilon_{\chi}^{X} &= \frac{S}{\rho}\int d\Pi_a d\Pi_b d\Pi_1 d\Pi_2 d\Pi_3 f_a f_b (E_1 + E_2)\\
        &\times (2\pi)^4 \delta^4(P_a + P_b - P_1 - P_2 - P_3) |\mathcal{M}|^2. \,
    \end{split}
\end{equation}

Neglecting the recoil of the target particle enables us to set $E_1 + E_2 = E_a$. Given these assumptions, and including the effects of Debye-screening, the square matrix element for the Primakoff-like production of scalar dark matter pairs off target particle $X$ is~\cite{Raffelt:1996wa}
\begin{equation*}
    |\mathcal{M}|^2 \approx \frac{4 \alpha^3 Q_X^2 m_X^2}{\pi \Lambda_s^4}\frac{p_a^2(1-\cos^2\theta)}{q^2 + k_s^2} \,,
\end{equation*}
which is independent of $p_b$, $p_1$, and $p_2$. Here, $k_s$ is the Debye-H\"uckel wavenumber,
\begin{equation}
    k_s^2 = \frac{4\pi\alpha}{T}\sum_i Q_i^2 n_i \,,
\end{equation}
where the sum is over all species in the plasma, with number density $n_i$ and charge $Q_i$. Substituting this into \cref{eq: Primakoff general energy loss}, integrating, and summing over all targets in the plasma, we find
\begin{equation}
\begin{split}
    \epsilon_{\chi}^{\mathrm{P},s}=\frac{\alpha^2 T^9}{32\pi^5\Lambda_s^4\rho}& \frac{\kappa^2}{2\pi^2}\int_0^{\infty} \frac{dx\, x^5}{\exp(x)-1}\Bigg\{ \bigg(\frac{1}{2}-\frac{1}{3}\frac{\kappa^2}{x^2}\bigg)\\
    &+ \frac{\kappa}{3 x}\bigg(\frac{\kappa^3}{x^3}+\frac{3\kappa}{x}\bigg)\log\bigg(1+\frac{x^2}{\kappa^2}\bigg)\\
    &- \frac{4\kappa}{3x}\tan^{-1}\bigg(\frac{x}{\kappa}\bigg)\Bigg\} \,,
\end{split}
\end{equation}
where $\kappa=k_s/(2T)$. Given standard PNS conditions, with $\kappa^2=1.41$~\cite{Raffelt:1996wa}, we find $\epsilon_{\chi}^{\mathrm{P},s}$ to be approximately a factor of $23$ smaller than $\epsilon_{\chi}^{\mathrm{ann},s}$, and therefore neglect its contribution in our limits.

For fermionic dark matter, the square matrix element for Primakoff like dark matter production can be expressed as
\begin{equation}
|\mathcal{M}|^2_{\rm{P}}=|\mathcal{M}|^2_{s}\bigg[\frac{m_{12}^2}{2} - 2m_{\chi}^2\bigg] \,,
\end{equation}
when averaged over all spins and polarisations, where $m_{12}$ is the invariant mass of the dark matter pair. With the approximations above, this simplifies to
\begin{equation}
|\mathcal{M}|^2=\frac{2\alpha^3 Q_X^2 m_X^2}{\pi\Lambda_f^6}\frac{p_a^2q(1-\cos^2\theta)}{q^2+k_s^2}\bigg(p_a \cos\theta - \frac{q}{2}\bigg) \,.
\end{equation}
This corresponds to the energy-loss rate
\begin{equation}
    \begin{split}
            \epsilon_{\chi}^{\mathrm{P}, f} &=\frac{\alpha^2 T^{11}}{8\pi^5\Lambda_f^6 \rho}\frac{\kappa^2}{2\pi^2}\int_0^{\infty}\frac{dx\, x^7}{\exp(x)-1}\Bigg\{ \bigg(\frac{1}{18}+\frac{7}{4}\frac{\kappa^2}{x^2} - \frac{1}{6}\frac{\kappa^4}{x^4}\bigg)\\
            &+ \frac{\kappa^2}{x^2}\bigg(\frac{1}{6}\frac{\kappa^4}{x^4}+\frac{\kappa^2}{x^2}-\frac{1}{2}\bigg)\log\bigg(1+\frac{x^2}{\kappa^2}\bigg)\\
            &- \frac{8\kappa^3}{3 x^3}\tan^{-1}\bigg(\frac{x}{\kappa}\bigg)\Bigg\} \,,
    \end{split}
\end{equation}
which is smaller than the equivalent expression from photon annihilation by a factor of approximately 200, given standard PNS conditions, and is therefore negligible.

\subsection{Nucleon bremsstrahlung}
\label{app: Nucleon brems}

Models of the SN1987A proto-neutron star suggest that its nucleon number density was larger than that of photons by upwards of two orders of magnitude~\cite{Nakazato:2012qf, Fischer:2016cyd}. Pair production via nucleon bremsstrahlung ($N N \rightarrow N N \chi \chi$) thus has the potential to rival photon annihilation, despite being suppressed by an additional factor of $\alpha^2$. However, the lowest order contribution to the dark-matter--nucleon bremsstrahlung matrix element cancels when all diagrams are included~\cite{Dev:2020eam, Hardy:2024gwy}. The energy-loss rate is therefore lower than na\"ively expected by a factor of $T^2 / m_N^2$, which renders it inconsequential compared with photon annihilation. For completeness, we briefly derive the energy loss rate below.

The energy-loss rate per unit mass to dark matter production via nucleon bremsstrahlung, $N (P_a) + N (P_b)\rightarrow N (P_1) + N (P_2) + \chi (P_3) + \chi (P_4)$, is given by
\begin{equation}
\label{eq: Brem E-loss general}
\begin{split}
    \epsilon_{\chi}^{\rm{brem}} &= \frac{S}{\rho}\int d\Pi_a d\Pi_b d\Pi_1 d\Pi_2 d\Pi_3 d\Pi_4 (2\pi)^4\\
    &\times \delta^4(E_a + E_b - E_1 - E_2 - E_3 - E_4)\\
    &\times \Big(\sum_{\rm{spins}}|\mathcal{M}|^2\Big) (E_3 + E_4) f_a f_b \,,
\end{split}.
\end{equation}
This expression can be evaluated in a similar manner to the expression for the production of a CP-even scalar via nucleon bremsstrahlung in Ref.~\cite{Dev:2020eam}. By defining,
\begin{equation}
\begin{split}
    \mathbf{p}_a &= \mathbf{P} + \mathbf{p}_i, \quad\mathbf{p}_b = \mathbf{P} - \mathbf{p}_i \,, \\
    \mathbf{p}_1 &= \mathbf{P} + \mathbf{p}_f, \quad\mathbf{p}_2 = \mathbf{P} - \mathbf{p}_f \,,
\end{split}
\end{equation}
and introducing the dimensionless parameters
\begin{equation}
    \begin{split}
        u\equiv \frac{p_i^2}{m_N T},\quad v&\equiv \frac{p_f^2}{m_N T},\quad z\equiv \cos\theta_{if} \,, \\
        \quad q\equiv \frac{m_{\chi}}{T},\quad x_3&\equiv \frac{E_3}{T},\quad x_4\equiv \frac{E_4}{T} \,, \\
        dz_{34}&\equiv\cos\theta_{34} \,,
    \end{split}
\end{equation}
\cref{eq: Brem E-loss general} can be simplified to
\begin{equation}
\label{eq: Brem E-loss simplified}
    \begin{split}
        \epsilon_{\chi}^{\mathrm{brem}} &= \frac{n_B^2 T^5}{2^{15}\pi^{11/2} m_N^2 \rho}\bigg(\frac{T}{m_N}\bigg)^{1/2}\int_{2q}^{\infty}du\int_{0}^{u-2q}dv \\
        &\int_{-1}^{+1}dz\int_{q}^{u-v}dx_3\int_{q}^{u-v-q}dx_4\int_{-1}^{+1}dz_{34} \\
        &e^{-u}\sqrt{uv}(u-v)\sqrt{x_3^2-q^2}\sqrt{x_4^2-q^2} \\
        &\Big(\sum_{\rm{spins}}|\mathcal{M}|^2\Big) \delta(u-v-x_3-x_4) \,.
    \end{split}
\end{equation}
The square matrix element for this process is identical to that for the emission of a single scalar in Ref.~\cite{Hardy:2024gwy} with the replacement of the scalar mass and energy by the invariant mass of the dark matter pair, $m_{34}$, and their total energy, $E_{34}=E_3 + E_4$. This gives
\begin{equation}
    \label{eq: brem ME s}
    \sum_{\rm{spins}}|\mathcal{M}|_{s}^2=1024\pi^2 \frac{g_{\chi N}^2}{m_N^2} \frac{m_{34}^4}{E_{34}^4} E_{\rm{CM}}^2 \bigg(\frac{d\sigma_{N}}{d\Omega}\bigg)
\end{equation}
for scalar dark matter and
\begin{equation}
    \label{eq: brem ME f}
    \sum_{\rm{spins}}|\mathcal{M}|_{f}^2=2\big[m_{34}^2-4m_{\chi}^2\big] \sum_{\rm{spins}}|\mathcal{M}|_s^2
\end{equation}
for fermionic dark matter, where $E_{\rm{CM}}\approx 2m_N$ is the centre-of-mass energy for the process, and $(d\sigma_{N}/d\Omega)$ is the differential cross-section of $2\rightarrow2$ nucleon scattering in the centre of mass frame. We can evaluate this expression by integrating over the solid angle and substituting empirical values for $\sigma_{N}$, e.g.\ those from Ref.~\cite{Norbury2013NucleonNucleonTC}. Considering this energy-loss channel only and applying the Raffelt criterion excludes $\Lambda_s\lesssim 4.86$\,GeV and $\Lambda_f\lesssim 1.68$\,GeV, far below the limits from photon annihilation.

\end{document}